\documentclass
[%
preprint,
 showpacs,preprintnumbers,
 amsmath,amssymb,
 aps,
 pre,
 floatfix,
 longbibliography
]{revtex4-1}

\usepackage{booktabs}
\usepackage{array}
\newcolumntype{L}[1]{>{\raggedleft\arraybackslash}p{#1}}
\AtBeginDocument{
\heavyrulewidth=.08em
\lightrulewidth=.05em
\cmidrulewidth=.03em
\belowrulesep=.65ex
\belowbottomsep=0pt
\aboverulesep=.4ex
\abovetopsep=0pt
\cmidrulesep=\doublerulesep
\cmidrulekern=.5em
\defaultaddspace=.5em
}

\usepackage{color}
\usepackage{xcolor}
\usepackage{graphicx}
\usepackage{dcolumn}
\usepackage{bm}
\usepackage{cancel}
\usepackage[utf8]{inputenc}

\begin{document}


\title{The role of the thermodynamic factor on the intrinsic and tracer diffusivities in binary mixtures}

\author{M. Sampayo Puelles}
\author{M. Hoyuelos}
\email{hoyuelos@mdp.edu.ar}

\affiliation{Instituto de Investigaciones F\'isicas de Mar del Plata (IFIMAR -- CONICET), Departamento de F\'isica, Facultad de Ciencias Exactas y Naturales,
Universidad Nacional de Mar del Plata, De\'an Funes 3350, 7600 Mar del Plata, Argentina}

\date{\today}

\begin{abstract}
One of the Darken equations gives a relationship between the intrinsic and the tracer diffusion coefficients, $D_A$ and $D_A^*$, of species $A$ in a solid binary mixture. In its original formulation, the equation reads $D_A = D_A^* \Gamma$, with $\Gamma$ the thermodynamic factor. The question addressed in this paper is how $D_A$ and $D_A^*$ depend separately on $\Gamma$. Using a recent result for transition probabilities in terms of the excess chemical potential (M. Di Muro and M. Hoyuelos, Phys. Rev. E \textbf{104}, 044104, 2021), it is shown that the intrinsic diffusivity does not depend on $\Gamma$. This approach simplifies a previous theoretical analysis that reaches the same result. Experimental results of diffusion in a Ni-Pd and Fe-Pd alloys (M. J. H. van Dal \textit{et al.}, Acta mater. \textbf{48}, 385, 2000) are used to check the theory. Numerical simulations of Ni-Pd were performed to show that the migration energy is the main factor responsible for the increase in diffusivity at intermediate concentrations.
\end{abstract}


\maketitle

\section{Introduction}

In 1948, Darken \cite{darken} obtained two equations to describe substitutional diffusion in solid binary mixtures, governed primarily by the presence of vacancies. Let us consider a binary mixture composed by species $A$ and $B$ that have molar concentrations $c_A$ and $c_B$.  The Darken equations represented a major advancement in the theoretical understanding of diffusion processes by establishing a connection between diffusivity and the thermodynamic factor, defined as $\Gamma = \beta \frac{\partial \mu_A}{\partial \log c_A}$, where $\beta=(k_B T)^{-1}$, and $\mu_A$ is the chemical potential (per particle) of species $A$. In terms of the excess chemical potential, the thermodynamic factor is $\Gamma = 1 + \beta \frac{\partial \mu_\text{ex}^A}{\partial \log c_A}$. It can be shown, through the Gibbs-Duhem relationship, that the thermodynamic factor is the same for both species.

The diffusion current (moles per unit area and time) for species $A$ with respect to the crystalline lattice, along the $x$ axis, is given by 
\begin{equation}\label{e.JA}
j_A = -D_A \frac{\partial c_A}{\partial x},
\end{equation}
where $D_A$ is the intrinsic diffusion coefficient for species $A$, different, in general, from $D_B$; this difference gives rise to a volume flux through a plane of the lattice perpendicular to the current direction. In the laboratory reference frame, where volume flux is zero, $A$ and $B$ have the same diffusion coefficient, $\tilde{D}$, known as the interdiffusion coefficient and given by
\begin{equation}\label{e.interdiff}
\tilde{D} = v_B c_B D_A + v_A c_A D_B,
\end{equation}
where $v_A$ and $v_B$ are the partial molar volumes; see, for example, \cite{mehrer,paul,shewmon}.

On the other hand, the diffusivity of a tagged particle of species $A$ in the mixture is given by the tracer diffusion coefficient $D_A^*$, connected with the mobility, $B_A$, through the Einstein relation, $D_A^* = B_A R T$, where $R$ is the ideal gas constant. Darken showed that $D_A$ and $D_A^*$ are related through the thermodynamic factor. More specifically, 
\begin{equation}\label{e.dark1}
D_A = D_A^* \frac{v_m}{v_B}\Gamma,
\end{equation}
where $v_m = N_A v_A + N_B v_B$ is the total molar volume, with $N_A$ and $N_B$ the mole fractions. Originally, Darken considered species with similar molar volumes such that $v_m\simeq v_A\simeq v_B$, and $D_A \simeq D_A^* \Gamma$. Combining Eqs.\ \eqref{e.interdiff} and \eqref{e.dark1}, the interdiffusion coefficient can be written as
\begin{equation}\label{e.dark2}
\tilde{D} = (N_A D_B^* + N_B D_A^*) \Gamma.
\end{equation}
Eqs.\ \eqref{e.dark1} and \eqref{e.dark2} are known as Darken equations.

Information about how $D_A$ and $D_A^*$ separately depend on the thermodynamic factor was obtained in Ref.\ \cite{dipietro3}. It was shown that the tracer diffusivity, $D_A^*$, behaves as $1/\Gamma$ and that the intrinsic diffusivity, $D_A$, does not depend on $\Gamma$. In this paper, we present an alternative and simpler derivation of the same results using an expression for transition rates in terms of the excess chemical potential that was recently derived in \cite{dimuro}. 

The paper is organized as follows. The form of transition rates is introduced in Sec.\ \ref{s.transition}. In Sec.\ \ref{s.intrinsic}, it is demonstrated that the intrinsic diffusion coefficient does not depend on the thermodynamic factor. Numerical evaluation of the Debye frequency for the Ni-Pd alloy are also included in Sec.\ \ref{s.intrinsic} in order to justify a linear approximation in its concentration dependence. Verification of the theoretical results using experimental data of the intrinsic diffusivity in solid mixtures of Ni-Pd and Fe-Pd, taken from Ref.\ \cite{vandal2}, is presented in Sec.\ \ref{s.compar}. The question of whether vacancy or migration energy is responsible of the observed increase in diffusivity at intermediate concentrations is addressed in Sec.\ \ref{s.vacmig}. Conclusions are presented in Sec.\ \ref{s.conclusions}.

\section{Transition rates}
\label{s.transition}

When vacancies and atoms, of species $A$ and $B$, occupy sites of the same lattice we have a substitutional alloy. Movement of atoms through vacancies is the dominant diffusion mechanism in substitutional alloys. This is a frequent situation when atoms are of similar size. An atom has to overcome a migration energy in order to abandon its position in the lattice and, simultaneously, a vacancy should be present in the destination site. Then, the jump rate of an atom is characterized by two energies: the migration energy (of species $A$), $G_M^A$, and the vacancy formation energy, $G_V$. The combination of both is the activation energy: $G_A = G_M^A + G_V$ for species $A$. The jump rate for an atom of species $A$ is 
\begin{equation}\label{e.WA}
W_A = \omega_A e^{-\beta G_A},
\end{equation}
where $\omega_A$ is the jump attempt frequency, of the order of the Debye frequency (see, for example, \cite[Sec.\ 5.3.5]{paul}).

Eq.\ \eqref{e.WA} is based on a picture at the microscopic level, since jumps between neighboring lattice sites are taken into account. Interactions at a thermodynamic level, represented by the excess chemical potential, are not explicitly represented in \eqref{e.WA}.

The thermodynamic aspects of transition rates and diffusivity are analyzed in Ref.\ \cite{dimuro}. A coarse grained picture is adopted in which microscopic details are lost. The system is divided into cells; each cell of size $a$ has volume $V=a^3$ and contains many lattice sites. Two neighboring cells, labeled 1 and 2, contain $n_1$ and $n_2$ atoms of species $A$ respectively. It can be demonstrated that the transition rate per particle between the two cells depends on the excess chemical potential in the following way \cite{dimuro}:
\begin{equation}\label{e.Wnn}
W^A_{n_1,n_2} = \nu_A  \frac{e^{-\beta \mu_{\text{ex},n_1}^A/2}}{\Gamma_{n_1}^{1/2}} \frac{e^{\beta \mu_{\text{ex},n_2}^A/2}}{\Gamma_{n_2}^{1/2}},
\end{equation}
where the excess chemical potential and the thermodynamic factor with sub-index $n_i$ are evaluated at particle concentration $n_i/V$, and $\nu_A$ is a jump frequency, independent of the excess chemical potential; $\nu_A$ contains information of the substratum, such as the number of vacancies, that is not included in this coarse grained picture. The order of sub-indices in $W^A_{n_1,n_2}$ indicates the jump direction, from cell 1 to cell 2. The derivation of \eqref{e.Wnn} is based on statistical mechanics concepts, such as the detailed balance relationship and the Widom insertion formula.

Eqs.\ \eqref{e.WA} and \eqref{e.Wnn} contain information at different levels. The main difference between them is that \eqref{e.WA} is based on a microscopic picture of jumps between neighboring lattice sites, while jumps between cells containing many lattice sites are considered in the derivation of \eqref{e.Wnn}. This last approach allows an explicit representation of the transition rate dependence on the excess chemical potential. Both equations are used in the next section to obtain the intrinsic diffusion coefficient.


\section{Intrinsic diffusion coefficient}
\label{s.intrinsic}

The purpose of this section is to determine the dependence of the intrinsic diffusion coefficient, $D_A$, on the thermodynamic factor and on concentration. The transition rates \eqref{e.Wnn}, that contain the thermodynamic information, can be used to calculate the particle current. Smooth spatial variations of the concentration are assumed. The average particle concentration is $\rho=\bar{n}/V$, and concentrations in each cell are $\rho_1=n_1/V$ and $\rho_2=n_2/V$, with $n_1\simeq n_2 \simeq \bar{n}$. Let us assume that cells 1 and 2 are aligned along the $x$ axis. The number of particles per unit time that jump between cells is $n_1 W_{n_1,n_2} - n_2 W_{n_2,n_1}$ and the area connecting cells is $a^2$. Then, the particle current is
\begin{align}
J_A &= (n_1 W_{n_1,n_2}^A - n_2 W_{n_2,n_1}^A)/a^2 \nonumber \\
&= \frac{\nu_A}{a ^2(\Gamma_{n_1}\Gamma_{n_2})^{1/2}}(n_1 e^{-\beta \Delta \mu_{\text{ex}}^A/2} - n_2 e^{\beta \Delta \mu_{\text{ex}}^A/2}) \nonumber \\
&\simeq \frac{\nu_A}{a^2 \Gamma} [n_1 - n_2 - \beta (n_1+n_2)\Delta \mu_{\text{ex}}^A/2] \nonumber \\
&\simeq - \frac{\nu_A}{a^2 \Gamma}\Delta n \underbrace{\left[1 + \beta \bar{n} \frac{\Delta\mu_\text{ex}^A}{\Delta n}\right]}_\Gamma = -\frac{\nu_A}{a^2}\, \Delta n \nonumber\\
&= - \nu_A a^2 \frac{\Delta \rho}{a},
\label{eq:J}
\end{align}
where $\Delta n = n_2-n_1$ and $\Delta \mu^A_{\text{ex}} = \mu^A_{\text{ex},n_2}-\mu^A_{\text{ex},n_1}$. The particle current is proportional to the concentration gradient, $\Delta \rho/a$, that is, the first Fick's law, and the proportionality constant is the intrinsic diffusion coefficient:
\begin{equation}\label{e.intr}
D_A = \nu_A a^2.
\end{equation}
The resulting coefficient is independent of the thermodynamic factor, or the excess chemical potential. This information is used below to support a simple approximation for the intrinsic diffusivity.


From Eq.\ \eqref{e.WA} we have
\begin{equation}\label{e.intr2}
D_A \propto e^{-\beta G_A + \ln \omega},
\end{equation}
where $\omega$ is the Debye frequency, a function of the mole fraction $N_A$. The proportionality can be written in terms of the diffusivity, the activation energy and the Debye frequency in the limit of small concentration, $D_{A0}$, $G_{A0}$ and $\omega_0$:
\begin{equation}\label{e.intr3}
D_A = D_{A0} e^{-\beta (G_A-G_{A0}) + \ln (\omega/\omega_0)}.
\end{equation}
(In the limit of small concentration of species $A$, interactions between $A$ atoms can be neglected and we have that $D_A = D_A^* = D_{A0}$.)

Then, knowing that $D_A$ does not depend on the excess chemical potential reduces the problem of determining the concentration dependence of the intrinsic diffusivity to obtaining the concentration dependence of the activation energy $G_A$ and the Debye frequency $\omega$. 

\begin{figure}
	\includegraphics[width=\linewidth]{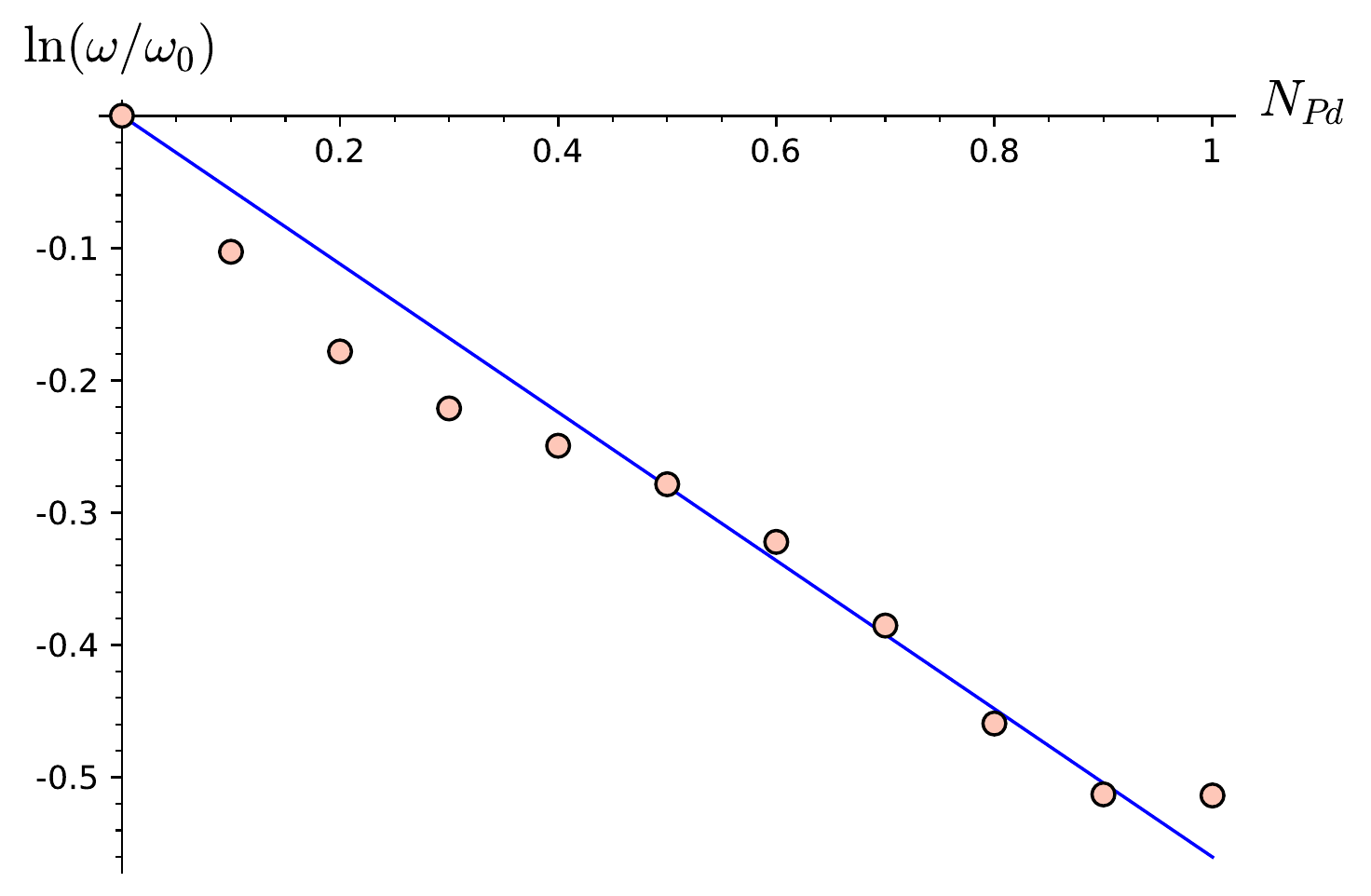}
	\caption{Numerical results of the logarithm of the relative Debye frequency $\omega/\omega_0$ against Pd mole fraction, $N_{\rm Pd}$, in a solid mixture Ni-Pd at zero temperature. The linear fit captures the approximately linear behavior. See Sec.\ \ref{s.debyemethod} for the simulation details.}
	\label{f.w}
\end{figure}

Molecular dynamics simulations were performed to obtain the Debye frequency for different mole fractions in a solid binary mixture of Ni-Pd at zero temperature; see Sec.\ \ref{s.debyemethod} for the methodology. The results are shown in Fig.\ \ref{f.w}; $\ln(\omega/\omega_0)$ has an approximately linear behavior against mole fraction. Some variation of the Debye frequency with temperature is expected; however it is assumed that the linear behavior is preserved for different temperatures, so that
\begin{equation}\label{e.lnw}
\ln (\omega/\omega_0) \simeq c N_A,
\end{equation} 
where $c$ is a constant.

The two extreme values of the activation energy are $G_{A0}$, for $N_A=0$, and $G_{A1}$, for $N_A=1$. A simple expression for $G_A$ based on the Vegard's law is proposed. The activation energy is approximated by
\begin{equation}\label{e.GA}
G_A = N_A G_{A1} + N_B G_{A0} - \varepsilon_A N_A N_B,
\end{equation}
where the first two terms correspond to the Vegard's law, a linear approximation between the two extreme values of $G_A$, and the last term is a possible deviation including the next non-linear term in the molar fraction. 

Replacing \eqref{e.lnw} and \eqref{e.GA} in \eqref{e.intr3}, and knowing that the value of the intrinsic diffusivity for the pure system is $D_{A1}= D_{A0} e^{-\beta (G_{A1}-G_{A0})+c}$, we have
\begin{equation}\label{e.intr4}
D_A = D_{A0}^{N_B} \, D_{A1}^{N_A}\, e^{\beta \varepsilon_A N_A N_B}.
\end{equation}

Using the Darken equation \eqref{e.dark1} combined with \eqref{e.intr3}, the tracer diffusivity is
\begin{equation}\label{e.tracer}
D_A^* = \frac{v_B}{v_m} \frac{1}{\Gamma} D_{A0} e^{-\beta (G_A-G_{A0})+ \ln (\omega/\omega_0)}.
\end{equation}
We obtained that $D_A^*$ behaves as $1/\Gamma$. A concentration of vacancies in thermal equilibrium is assumed in the derivation of the Darken equation. This is an approximation that works in many cases, but it is not always valid. Vacancies are created and annihilated at opposite sites of the interdiffusion zone due to the volume flux (Kirkendall effect) that was mentioned in the introduction. The Darken-Manning equations, that consider the effect of vacancy-wind factors, include necessary corrections; see, for example, \cite[Sec.\ 10.4]{mehrer}. In order to avoid these difficulties, the comparison with experimental results presented in Sec.\ \ref{s.compar} is restricted to the intrinsic diffusion coefficient.

\subsection{Method to calculate the Debye frequency}
\label{s.debyemethod}

The Debye frequency is given by
\begin{equation}\label{e.debye}
\omega = b \frac{v_s}{\alpha}
\end{equation}
where $v_s$ is the sound speed, $\alpha$ is the lattice spacing and $b$ is a proportionality constant, equal to $(6\pi^2)^{1/3}$ for a cubic crystal \cite{kittel}. The sound in a solid has longitudinal and transverse modes, each one with speeds given by
\begin{align}
v_{L} &= \sqrt{\frac{K+4G/3}{\rho}} \\
v_{T} &= \sqrt{\frac{G}{\rho}},
\end{align}
where $K$ is the bulk modulus, $G$ is the shear modulus, and $\rho$ is the density \cite{kinsler}. Following Ref.\ \cite{holland}, we use the average sound speed given by
\begin{equation}\label{e.vs}
v_s = 3/(2/v_T + 1/v_L).
\end{equation}
The elastic constants, $K$ and $G$, were calculated using LAMMPS software \cite{plimpton}. A box of $6\times 6 \times 6$ unit cells of the Ni-Pd fcc lattice, with periodic boundary conditions, was considered. The box is deformed in different directions and the elastic constants are obtained from the change in the stress tensor. The procedure was repeated for different values of the mole fraction $N_{\rm Pd}$ from 0 to 1, using an average lattice spacing $\alpha$ that varies linearly from 3.521 Å (pure Ni) to 3.889 Å (pure Pd); the approximately linear behavior of $\alpha$ in Ni-Pd alloys was reported in \cite{bidwell}. The resulting value of $\omega$ was averaged over 10  samples for each value of the mole fraction. The interaction potential in the Ni-Pd alloy was taken from the Interatomic Potentials Repository \cite{ipr}; it is an angular-dependent potential of the Ni-Pd system, obtained by fitting the experimental data and first-principles calculations, reported in Ref.\ \cite{XuY}.

\section{Comparison with experiments}
\label{s.compar}

Eq.\ \eqref{e.intr4} for the intrinsic diffusivity was compared with experimental results in Ref.\ \cite{dipietro3} using data for the following mixtures: Au-Ni (at 900$^\circ$C) \cite{reynolds}, Ag-Au (at 894$^\circ$C) \cite{mead} and Fe-Pd (at 1150$^\circ$C) \cite{fillon}. Here we extend the experimental data set, including Ni-Pd (at 1100$^\circ$C) and Fe-Pd (at 1100$^\circ$C) \cite{vandal2}, to further test the validity of the equation.  The data used here for the mixture Fe-Pd, taken from Ref.\ \cite{vandal2}, were obtained using an experimental technique (diffusion couple technique including incremental and ``multi-foil'' couples) that allows a direct measurement of intrinsic diffusivity; instead, in Ref.\ \cite{fillon} the tracer diffusivity is measured and the intrinsic diffusivity is indirectly obtained from these measurements. 

\begin{figure}
	\includegraphics[width=\linewidth]{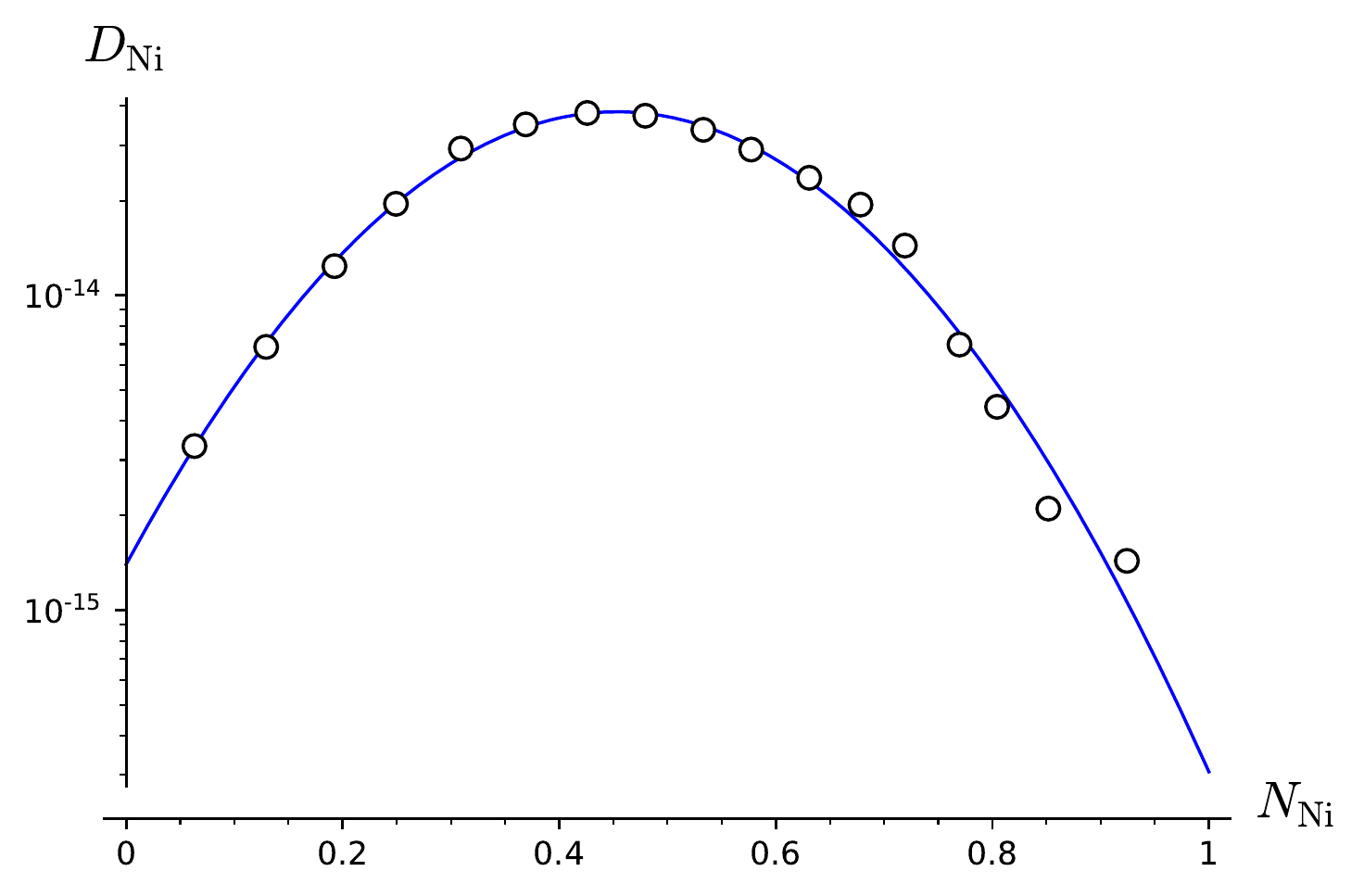}
	\includegraphics[width=\linewidth]{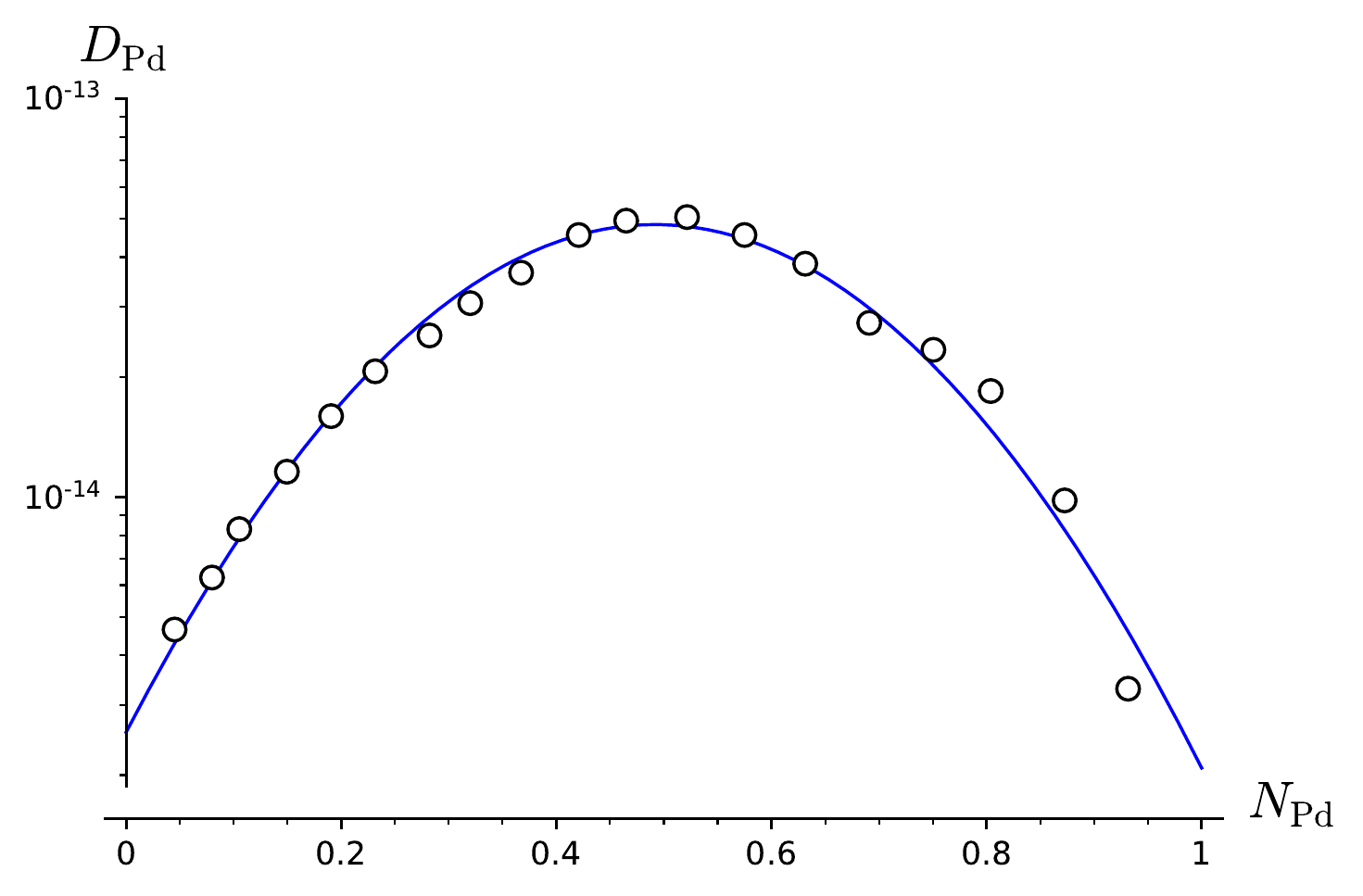}
	\caption{Intrinsic diffusivities, $D_\text{Ni}$ and $D_\text{Pd}$ (units: m$^2$/s), in the  Ni-Pd mixture against mole fraction. Circles correspond to experimental data of Ref.\ \cite{vandal2} and the curves correspond to Eq.\ \eqref{e.intr4} (see Table \ref{t.fitting} for the parameters used).}	\label{f.NiPd}
\end{figure}

\begin{figure}
	\includegraphics[width=\linewidth]{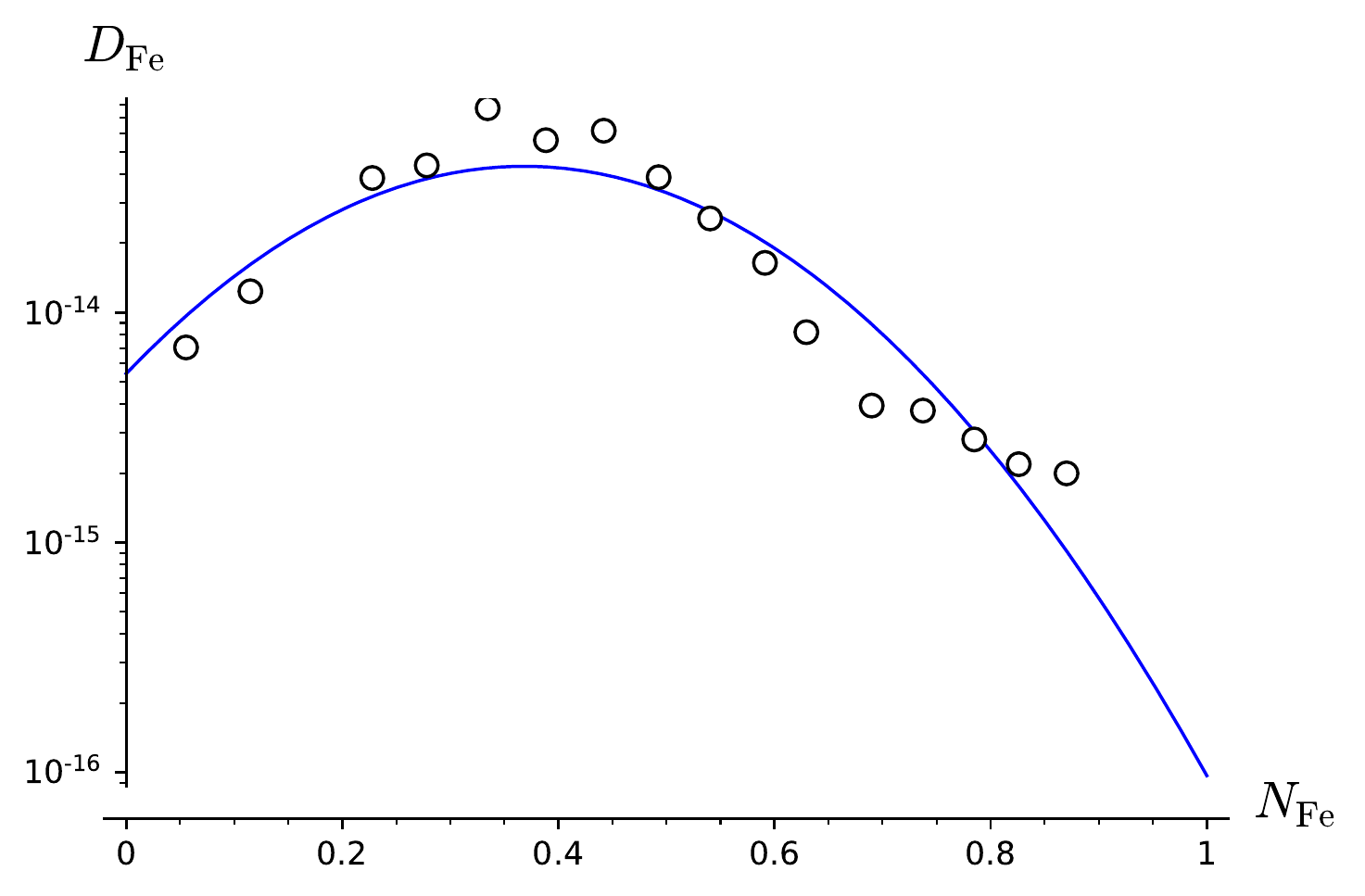}
	\includegraphics[width=\linewidth]{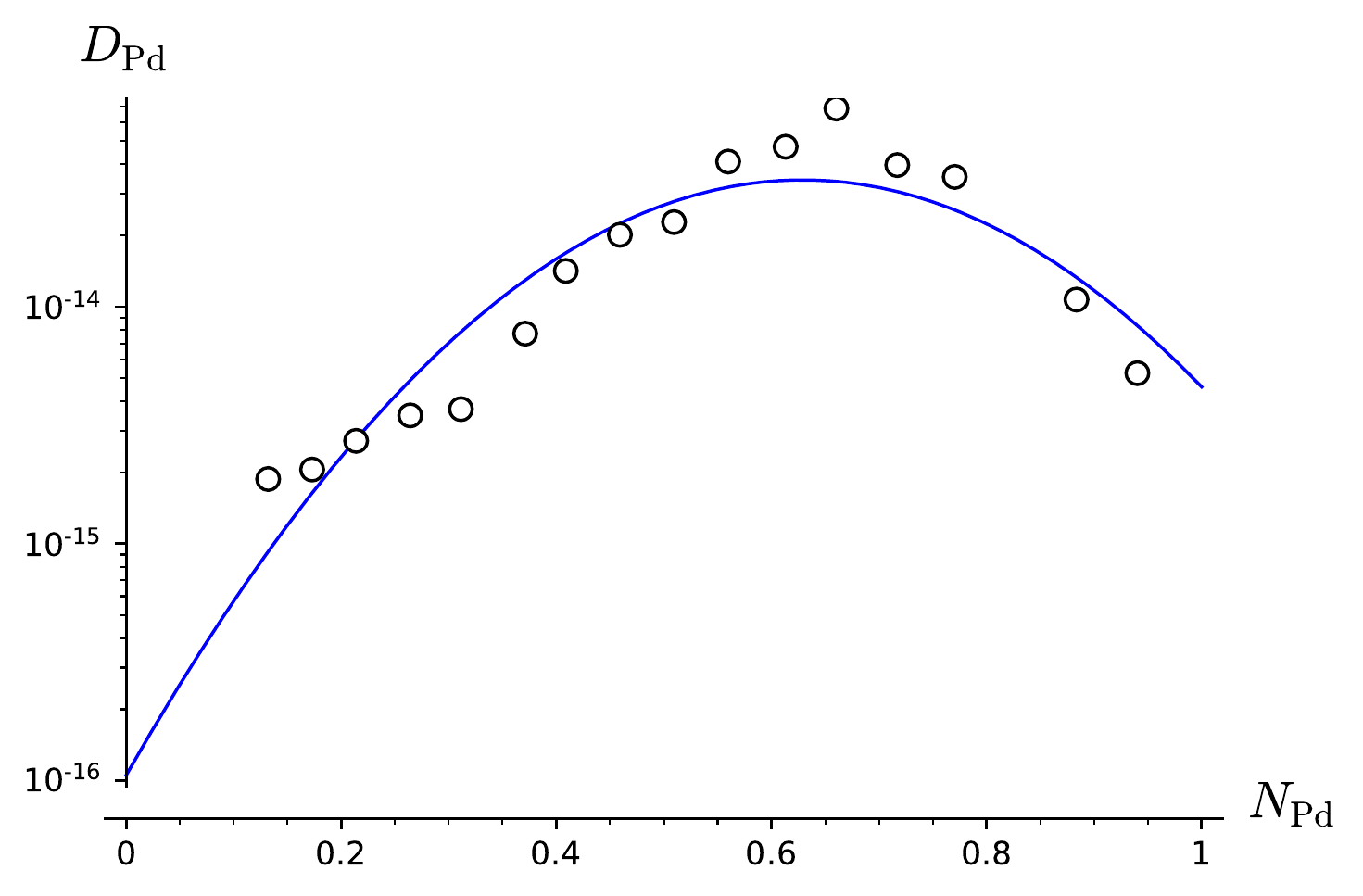}
	\caption{Intrinsic diffusivities, $D_\text{Fe}$ and $D_\text{Pd}$ (units: m$^2$/s), in the  Fe-Pd mixture against mole fraction. Circles correspond to experimental data of Ref.\ \cite{vandal2} and the curves correspond to Eq.\ \eqref{e.intr4} (see Table \ref{t.fitting} for the parameters used).}	\label{f.FePd}
\end{figure}

Figures \ref{f.NiPd} and \ref{f.FePd} show the intrinsic diffusivities for  Ni-Pd and Fe-Pd mixtures respectively, against the corresponding mole fractions. The curves correspond to Eq.\ \eqref{e.intr4} with adjusted values of $D_{0}$, $D_1$ and $\beta\varepsilon$ (sub-index $A$ is omitted for simplicity). It can be seen that the equation satisfactorily represents the data, specially for the Ni-Pd mixture. Table \ref{t.fitting} contains the values used for $D_{0}$, $D_1$ and $\beta\varepsilon$ in each case.

\begin{table}[]
	\centering
	\begin{tabular}{r|r|r|r}
		& $D_0$ & $D_1$ & $\beta\varepsilon$ \\
	\midrule
		Ni & 0.14  & 0.031  & 16.1  \\
		Pd & 0.26 & 0.21 & 12.2 \\ 
	\midrule
		Fe & 0.54 & 0.01 & 15.3 \\
		Pd & 0.01 & 0.46 & 14.6 \\
	\midrule
	\end{tabular}
	\caption{Adjusted parameters of Eq.\ \eqref{e.intr4} for each metal in their respective alloy. Units for $D_0$ and $D_1$ are $10^{-14}$m$^2$/s.}
	\label{t.fitting}
\end{table}

In the limit $N_A\rightarrow 1$, $D_{A1}$ is the self-diffusion coefficient of species $A$; it can be obtained from Table 13.1 in Ref.\ \cite{sohn}; the reference values for Ni, Pd and Fe at 1100$^\circ$C are $D_\text{Ni,1}=0.261$, $D_\text{Pd,1}=0.153$, and $D_\text{Fe,1}=0.06$, with units $10^{-14}$m$^2$/s. The differences with the values of Table \ref{t.fitting} can be understood as a consequence of the simple form assumed for the activation energy \eqref{e.GA}. Despite these discrepancies, the approximation for $G_A$ is still able to provide a good description of the intrinsic diffusivity in the whole range of the mole fraction.

\section{Vacancy versus migration energy}
\label{s.vacmig}

\begin{figure}
	\includegraphics[width=\linewidth]{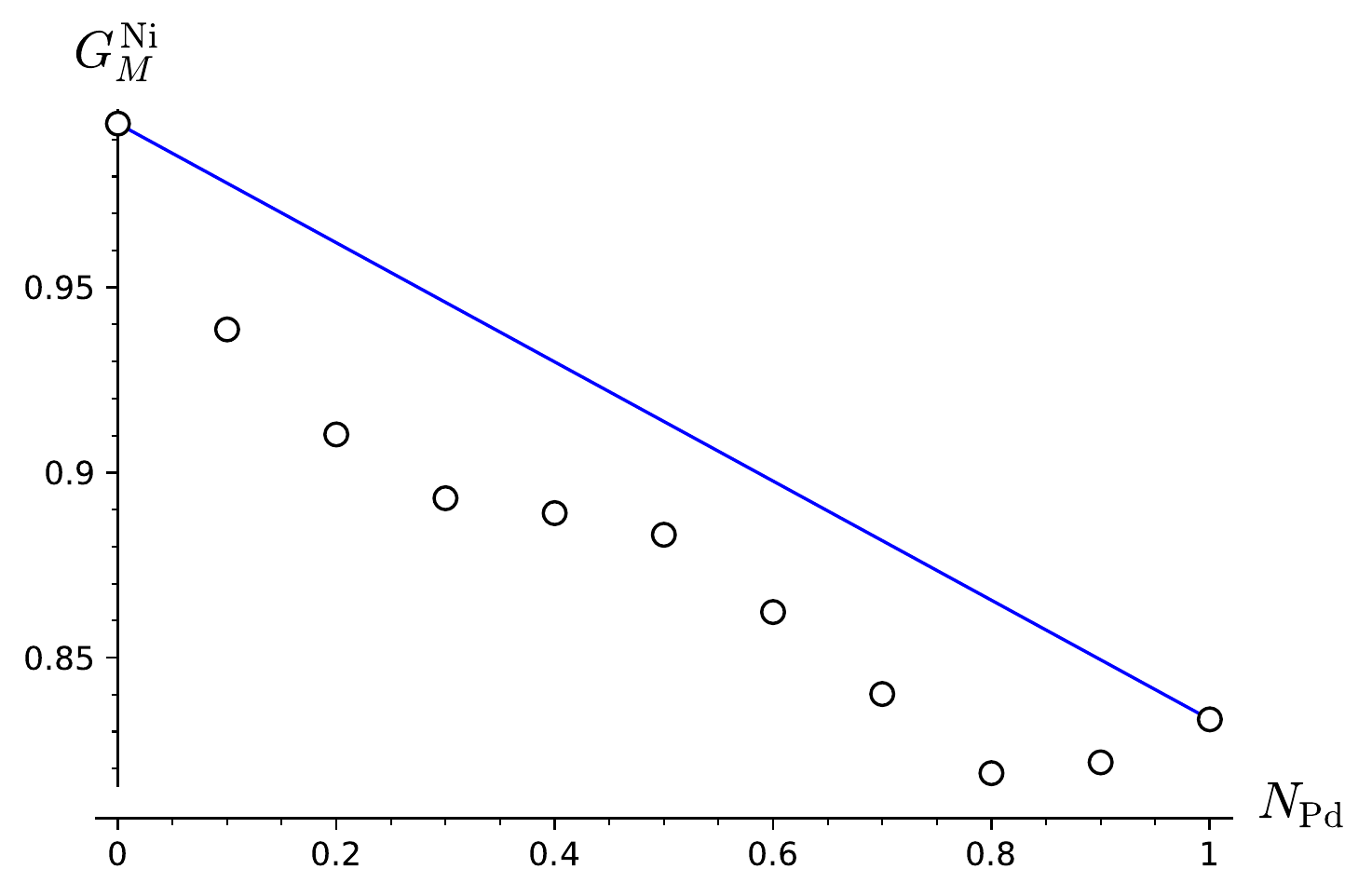}
	\includegraphics[width=\linewidth]{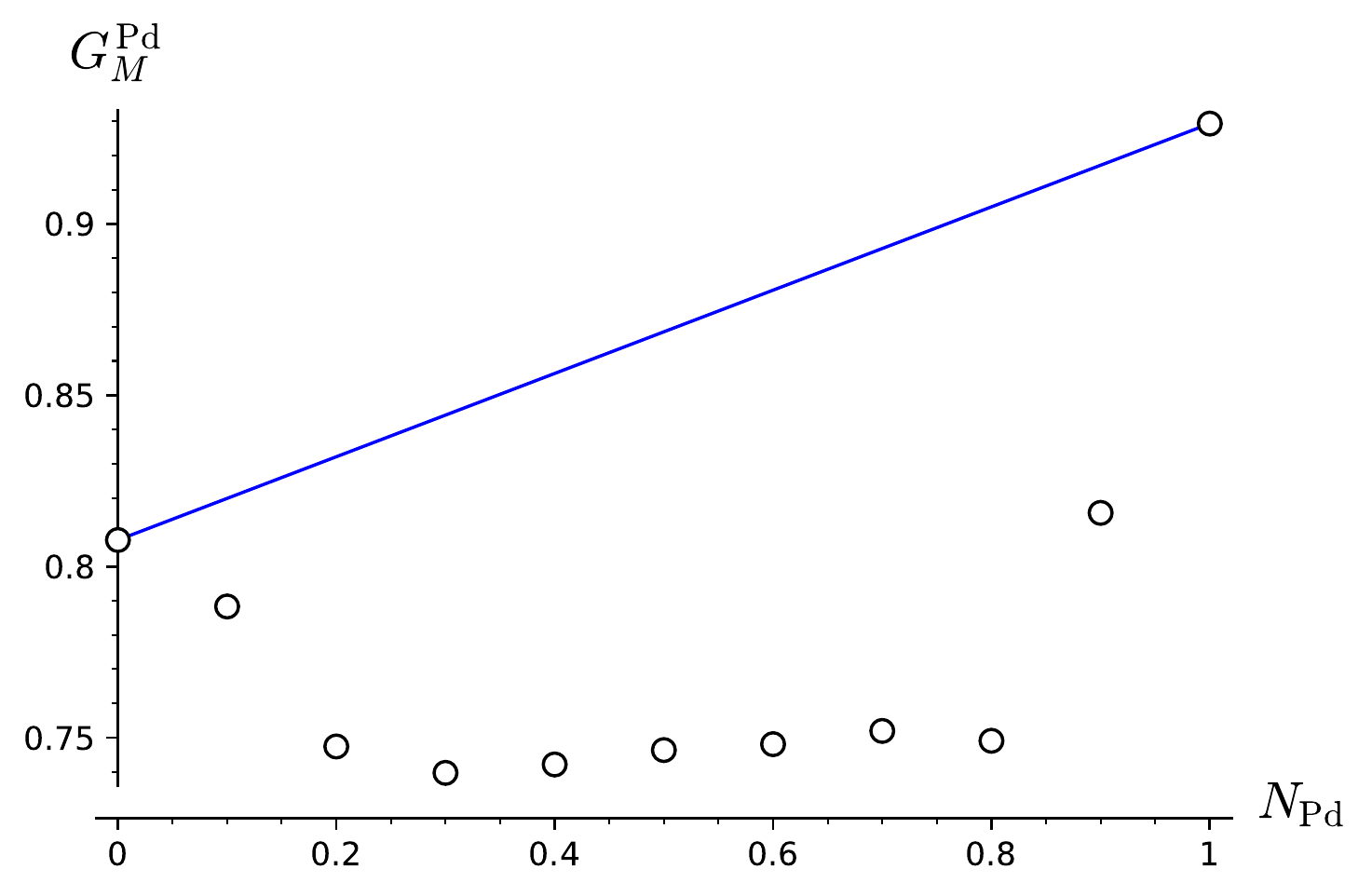}
	\caption{Numerical results of migration energy of Ni and Pd, $G_M^{\rm Ni}$ and $G_M^{\rm Pd}$ (units: eV), in a Ni-Pd solid mixture, against mole fraction, using the angle-dependent potential of Ref.\ \cite{XuY}. The line corresponds to the Vegard's law.  See Sec.\ \ref{s.migration} for the simulation details.}
	\label{f.mig}
\end{figure}

In all the mixtures analyzed, here and in Ref.\ \cite{dipietro3},  positive values of the non-linear term parameter, $\varepsilon_A$, were obtained. This means that the diffusivity is larger than the value predicted by Vegard's law for intermediate concentrations. The mixture enhances the diffusivity. The activation energy is the sum of the migration and the vacancy formation energies, $G_A = G_M^A + G_V$. It is interesting to establish whether the migration energy or the vacancy formation energy is the main responsible for the increase in diffusivity at intermediate concentrations. A possible interpretation of the results is that, for intermediate concentrations, the mixture of atoms with different sizes introduces a disorder in the lattice that favors the vacancy formation. But this is not actually the case. Numerical simulations show that the vacancy formation energy does not decrease in the mixture. On the contrary, in Ref.\ \cite{zhang} it was shown that an illustrative generic alloy system with ordering tendencies has a vacancy formation energy that increases at intermediate concentrations, meaning that the mean number of vacancies is smaller than the value predicted by Vegard's law; see Fig.\ 16 in \cite{zhang}. A qualitatively similar result was obtained in \cite{utt} for a high entropy alloy using grand-canonical lattice Monte Carlo simulations, in this case the vacancy formation energy is slightly above Vegard's law. These results indicate that the observed increase in diffusivity with respect to Vegard's law at intermediate concentrations is more likely a consequence of a decrease in migration energy than in vacancy formation energy. This supposition was checked with numerical simulations of the mixture Ni-Pd in which the migration energy was calculated; see Sec.\ \ref{s.migration} for the methodology. The  results obtained are shown in Fig.\ \ref{f.mig}; they present a decrease of the migration energy for intermediate concentrations.

\subsection{Method to calculate migration energy}
\label{s.migration}

The migration energy was calculated using LAMMPS software \cite{plimpton} and the nudged elastic band method (NEB). The NEB allows to find the height of an energy barrier associated with a transition state. In this case we use it for the transition of an atom towards a vacancy. 

As in Sec.\ \ref{s.debyemethod}, a box of $6\times6\times6$ unit cells of Ni-Pd fcc lattice, with periodic boundary conditions, was considered, and the interaction potential reported in Ref.\ \cite{XuY} was used; also, a linear behavior for the lattice spacing $\alpha$ as a function of mole fraction was assumed \cite{bidwell}.  A vacancy is created in the lattice and the species (Ni or Pd) of a first neighbor is set as the initial configuration, this atom is the one that will perform the hop. The final configuration corresponds to the vacancy at the site of the jumping atom and the atom at the site where the vacancy was initially found. 

For the hop of a Ni atom we considered 2000 different configuration of atoms in the mole fraction $N_{\rm Pd}$ range from 0.1 to 0.9. Note that for $N_{\rm Pd}$ equal to 0 or 1 it is not necessary to average different configurations, since all atoms are of the same type. For the case of Pd, 1000 realizations were enough.

We perform NEB calculation for six replicas, the first and last are the initial and the end point of the transition path. During the NEB calculation the set of replicas converge toward a minimum energy path of conformational states that transition over a barrier. The configuration of highest energy along the path correspond to a saddle point, and the potential energies for the set of replicas represents the energy profile of the transition along the minimum energy path.

Migration energy is the energy of the barrier, which is the difference in energy between the saddle point and the first replica. Both the final and initial replicas have approximately the same energy.

\section{Conclusions}
\label{s.conclusions}

The Darken equation \eqref{e.dark1} gives a relationship between the intrinsic and the tracer diffusion coefficients, $D_A$ and $D_A^*$, through the thermodynamic factor $\Gamma$. Nevertheless, it does not provide information about how $D_A$ and $D_A^*$ separately depend on $\Gamma$. This problem was addressed in Ref.\ \cite{dipietro3}, where it was shown that the intrinsic diffusivity does no depend on $\Gamma$. Here, we arrive at the same result using a general expression for transition rates that contains information at the thermodynamic level, that is, the expression determines how transition rates depend on the excess chemical potential \cite{dimuro}. A more direct derivation is obtained in this way; one important simplification is that it is not necessary to apply the concept of ``interpolation parameter'' used in \cite{dipietro3}. The procedure provides a deeper understanding of the problem of the dependence of diffusivity on the thermodynamic factor.

It is well known that vacancies play a fundamental role in substitutional diffusion. According to our result, the concentration dependence of the intrinsic diffusivity is completely determined by the Debye frequency and the form of the activation energy $G_A$, that includes the vacancy formation energy and the migration energy (as usual, experiments are more complicated than theoretical idealizations; some ingredients that are not taken into account, and that may be relevant, are, for example, the presence of impurities or the impurity vacancy binding energy; see \cite{santra}). Numerical simulations of the Ni-Pd alloy show that the Debye frequency behaves approximately linearly as a function of the mole fraction. A simple form for $G_A$ against mole fraction is proposed using the Vegard's law and including a quadratic term proportional to parameter $\varepsilon$. This approximation allows a theoretical description of the dependence of the intrinsic diffusivity on mole fraction, see Eq.\ \eqref{e.intr4}. Experimental data of diffusion in Ni-Pd and Fe-Pd mixtures \cite{vandal2} are consistent with the theoretical results. Positive values of $\varepsilon$ were obtained; this implies diffusion coefficient values that are larger than a linear interpolation between $D_{A0}$ and $D_{A1}$, for mole fractions 0 and 1 respectively. Regarding the question of whether this increase in diffusivity with respect to Vegard's law is mainly a consequence of a decrease in the migration energy or the vacancy formation energy, numerical simulation of other authors suggest that vacancy formation energy actually increases at intermediate concentrations. Therefore, the diffusivity increase at intermediate concentrations should be a consequence of a decrease of migration energy. Such decrease of the migration energy was numerically observed in the solid mixture of Ni-Pd.

Eq.\ \eqref{e.intr4} in logarithmic scale has the form of a parabola. A parabolic form for the intrinsic diffusion coefficient against concentration is assumed as an hypothesis in Refs.\ \cite{wierzba2,wierzba} where a method for calculating intrinsic diffusivities in multi component systems is proposed.

\begin{acknowledgments}
This work was partially supported by Consejo Nacional de Investigaciones Cient\'ificas y T\'ecnicas (CONICET, Argentina, PUE 22920200100016CO).
\end{acknowledgments}

\bibliography{mixtures.bib}

\end{document}